\documentclass[traditabstract]{aa}  
\usepackage{graphicx, xcolor}
\usepackage{longtable,lscape}
\usepackage{dcolumn}
\usepackage{txfonts}
\usepackage[authoryear]{natbib} 
\bibpunct{(}{)}{,}{a}{}{,} % citation like A&A style
\def\changed{\color{black}}
\def\changedtwo{\color{black}}

\begin{document}
   \title{The Quintuplet Cluster}

   \subtitle{II. Analysis of the WN stars~\thanks{Based on observations
   collected at the ESO VLT (program 077.D-0281(A)).}} 

   \author{A. Liermann \inst{1}
          \and
          W.-R. Hamann \inst{1}
          \and
          L. M. Oskinova \inst{1}
          \and
          H. Todt \inst{1}
          \and
          K. Butler
          \inst{2}
          }

   \institute{Universit\"at Potsdam, Institut f\"ur Physik und Astronomie,
              14476 Potsdam, Germany
%   \and
%              {\changed Max-Planck-Institut f\"ur Radioastronomie, 53121
%              Bonn, Germany}
   \and
              Universit\"ats-Sternwarte M\"unchen, 81679 M\"unchen, Germany}

   \date{Received June 10, 2009; accepted July 30, 2010}

   \titlerunning{The Quintuplet Cluster. II.}

  \abstract{ 

Based on $K$-band integral-field spectroscopy, we analyze four
Wolf-Rayet stars of the nitrogen sequence (WN) 
found in the inner part of the Quintuplet cluster. All WN stars
(WR\,102d, WR\,102i, WR\,102hb, and WR\,102ea) are of spectral subtype
WN9h. One further star, LHO\,110, is included in the analysis
{\changed which has been classified as Of/WN? previously but turns out
  to be most likely a WN9h star as well}. 
The  Potsdam Wolf-Rayet (PoWR) models for expanding atmospheres are
used to derive the fundamental stellar and wind parameters.

The stars turn out to be very luminous, $\log{(L/L_\odot)} > 6.0$,
with relatively low stellar temperatures, {\changed $T_* \approx$ 25--35\,kK}.
Their stellar winds contain a significant fraction of hydrogen, up to
{\changed $X_\mathrm{H} \sim 0.45$} (by mass). {\changed We discuss
the position of the Galactic center WN stars in the
Hertzsprung-Russell diagram and find that they form a distinct
group.} In this respect, the Quintuplet WN stars are similar to
late-type WN stars found in the Arches cluster and elsewhere in the Galaxy.

Comparison with stellar evolutionary models reveals that the
Quintuplet WN stars should have been initially more massive than
60\,$M_\odot$. They are about 2.4--3.6\,Million years old, and might
still be central hydrogen burning objects. The analysis of the
spectral energy distributions of the program stars results in a mean
extinction of $A_K = 3.1 \pm 0.5$\,mag ($A_V = 27 \pm 4$\,mag) towards
the \object{Quintuplet cluster}. 
} 


   \maketitle
%___________________________________________________________________

\section{Introduction}

The Galactic center (GC) is a unique region to study star formation
and evolution in the special environment close to the supermassive
black hole. Three very massive clusters have been found within 35\,pc
projected distance to the GC, hosting a large fraction of the known
Galactic massive stars such as luminous blue variables (LBV) and
Wolf-Rayet (WR) stars. {\changed The Central cluster is located within
1\,pc from the central black hole \citep{Krabbe+1991}. Two further
massive star clusters, the \object{Arches} and the
\object{Quintuplet}, are located in the vicinity of Sgr~A*.
Interestingly the three clusters differ in age, the 2.5\,Ma old
\object{Arches cluster} \citep{Figer+2002, Najarro+2004} contains many
OB and WN type stars, while the more evolved \object{Quintuplet
  cluster} and Central cluster with ages of about 4 and 6\,Ma
\citep{FMM99,Paumard+2006} contain WC stars as well. In addition, one
well known LBV, the Pistol star \citep{FMM99, FMG99}, and further LBV
candidates \citep{Geballe+2000, Barniske+2008} are found in the
vicinity of the \object{Quintuplet cluster}. 

Massive stars in general, and WR stars in particular, are important
sources of ionizing photons, momentum and chemical elements to impact
on the circumstellar medium.  Stars that display CNO-processed matter
in a strong stellar wind are classified as WR stars of the nitrogen
sequence (WN type). The cooler, late WN subtypes (WNL) usually contain
some rest of hydrogen in their atmospheres, while the hotter, early
subtypes (WNE) are hydrogen free \citep{Hamann+1991}. Typically, WNL
stars are significantly more luminous than WNE stars
\citep{HGL2006}. The WN stage may be followed by the WC stage, when
the products of helium burning appear in the stellar atmosphere.  Some
evolutionary scenarios suggest that the most massive of these stars
undergo LBV phases \citep{Langer+1994}, and the WNL phase might even
precede the LBV stage \citep{Crowther2007, Smith-Conti2008}, whereas
some evolutionary paths for stars with initial masses $>90\,M_\odot$
seem to skip the LBV stage completely \citep{Maeder+2008}. However,
the details of the evolution of massive stars are still under debate.

By means of quantitative spectral analysis in combination with
comprehensive stellar model atmospheres the stellar parameters can be
assessed. This allows to determine feedback parameters of the
population of massive stars in the Galactic center and will provide
constrains on stellar evolutionary models.

However, the high extinction in this direction ($A_{\rm V} \approx
28$\,mag) prohibits observations in the UV and optical
range. Therefore, IR spectroscopy is the prime tool to study massive
stars in the GC.  {\changedtwo Different ESO IR-instruments at La
Silla and VLT Paranal were used to observe the Central cluster
\citep{Najarro+1994,Krabbe+1995, Eckart+2004, Paumard+2006,
Martins+2007}, the \object{Arches cluster} \citep{Blum+2001, Martins+2008},
and the \object{Quintuplet cluster} \citep{Liermann+2009}.

\citet{Najarro+1997} and \citet{Martins+2007} both} applied CMFGEN
stellar atmosphere models \citep{Hillier-Miller1989} to analyze
$K$-band spectra of massive stars in the Central cluster. They found
that the stellar parameters of these stars are similar to other
Galactic WNL stars.  Most of the WNL stars studied by
\citet[][{\changedtwo note that some are classified only as Ofpe/WN9
stars}]{Martins+2007} appear relatively rich in hydrogen. Therefore,
the authors support the evolutionary sequence (Ofpe/WN9
$\leftrightarrow$ LBV) $\rightarrow$ WN8 $\rightarrow$ WN/C for most
of the observed stars.

In their study of the brightest stars in the \object{Arches cluster},
\citet{Martins+2008} found that these stars are either H-rich WN7-9
stars or O supergiants with an age of 2 to 4\,Ma. The WN7-9h stars
reveal high luminosities, $\log{L/L_\odot} = 6.3$, which is consistent
with initial stellar masses of $\sim$120$M_\odot$. The chemical
composition shows both N enhancement and C depletion along with a
still high amount of H, which leads the authors to conclude that the
stars are core H-burning objects.  It was found that the properties of
the Arches massive stars argue in favor of the evolutionary scenario
of \citet{Crowther-etal1995} O $\rightarrow$ Of $\rightarrow$ WNL +
abs $\rightarrow$ WN7.  Thus, the massive stars in the Central and in
the \object{Arches cluster} seem to conform to the standard evolutionary
models, and display stellar feedback parameters that do not
significantly differ from other massive stars in the Galaxy.

The present paper is based on our observations with the ESO-VLT
Spectrograph for INtegral Field Observation in the Near-Infrared
(SINFONI) of the central parts of the \object{Quintuplet cluster}. A spectral
catalog of the point sources (hereafter LHO catalog) was presented by
\citet{Liermann+2009}. The LHO catalog lists 13 WR stars, among them 4
WN and 9 WC types. In this paper we concentrate on the analysis of the
WN stars, while the analysis of WC stars will be subject of a
subsequent paper.} Additionally, we analyze the star \object{LHO\,110}, which
was classified as Of/WN candidate in the LHO catalog. However, the
spectrum of this star strongly resembles spectra of typical WN stars,
therefore we include \object{LHO\,110} in our sample.

In order to analyze our program stars (\object{WR\,102d},
\object{WR\,102i}, \object{WR\,102hb}, \object{WR\,102ea}, and
\object{LHO\,110}), we fit their spectra with the Potsdam 
Wolf-Rayet model atmospheres \citep[see also references
therein]{Hamann-Graefener2004} and derive the fundamental stellar
parameters.  {\changed By quantitative comparison of stellar
evolutionary models including rotation \citep{Meynet+Maeder2003}
with the stellar parameters the initial {\changed masses and ages}
of the sample stars are determined.  }

The paper is organized as follows. In Sect.\,\ref{sec:models} we give
a short characterization of the applied model atmospheres. The
spectral analyses is described in Sect.\,\ref{sec:analysis} and the
derived stellar parameters are discussed in
Sect.\,\ref{sec:discussion}. Therein, mass-loss rates from radio
free-free emission are debated (Sect.\,\ref{sec:radio}) and our
empirical results are compared with evolutionary models
(Sect.\,\ref{sec:evolution}). We conclude with a summary of the
obtained results in Sect.\,\ref{sec:conclusion}.

%__________________________________________________________________
\section{The models}
\label{sec:models}
The Potsdam Wolf-Rayet non-LTE model atmospheres
\citep[PoWR]{Hamann-Graefener2004} are based on the assumptions of
spherical symmetry and stationary mass loss. A velocity field is
pre-scribed by the standard $\beta$-law for the supersonic part, with
the terminal velocity $\varv_\infty$ as a free parameter. The exponent
$\beta$ is set to unity for all models. In the subsonic region the
velocity field is defined such that a hydrostatic density
stratification is approached.  Our models have the inner boundary set
to a Rosseland optical depth of 20, thus defining a ``stellar radius''
$R_*$ that represents the hydrostatic core of the stars. The ``stellar
temperature'' $T_*$ is given by the luminosity $L$ and the stellar
radius $R_*$ via the Stefan-Boltzmann law, i.e.\ $T_*$ denotes the
effective temperature referring to the radius $R_*$.

We account approximately for wind inhomogeneities (``clumping''),
assuming that optically thin clumps fill a volume fraction $f_{\rm V}$
while the interclump space is void. Thus, the matter density in the
clumps is higher by a factor $D = f_{\rm V}^{-1}$ compared to an
un-clumped model of same parameters. All models are calculated with
$D$\,=\,4 which is a rather conservative choice. Small scale random
motions are considered by applying a microturbulence velocity of
$\varv_{\rm D} = 100 \,{\rm km \, s^{-1}}$.  Model atoms comprise
hydrogen, helium, CNO and iron-group elements. The latter are
accounted for in a ``superlevel'' approach enabling blanketing by
millions of lines \citep{Graefener+2002}.

\section{The analysis}
\label{sec:analysis}

\subsection{{\changed The observations}}

We use flux-calibrated $K$-band spectra from the LHO catalog for the
stars \object{WR\,102d} (\object{LHO\,158}), \object{WR\,102i}
(\object{LHO\,99}), \object{WR\,102hb} (\object{LHO\,67}),
\object{WR\,102ea} (\object{LHO\,71}) and \object{LHO\,110}. {\changed
The data had been flux-calibrated by means of standards stars and then
applied for synthetic aperture photometry, yielding a mean error on
the obtained K-band magnitudes of 0.4\,mag (see LHO catalog for further
details).} For the comparison {\changed of the stellar parameters,}
we refer to previous studies of Galactic WN stars \citep{HGL2006} and
of the WN stars in the \object{Arches cluster} \citep{Martins+2008}. Two
further WR stars in the vicinity of the \object{Quintuplet cluster},
\object{WR\,102c} and \object{WR\,102ka}, have been analyzed by
\citet{Barniske+2008} recently. 

{\changed Our analysis is based on pronounced lines in the observed
$K$-band spectra, which are due to helium
and hydrogen: \ion{He}{i} at 2.059\,$\mu$m and 2.113\,$\mu$m,
\ion{He}{ii} at 2.189\,$\mu$m (10--7) , and Br$\gamma$ at 
2.166\,$\mu$m blended with \ion{He}{ii} (14--8) and various
\ion{He}{i} lines. A selection of $K$-band lines is compiled in
Table\,\ref{tab:Klinelist}.
}

%________________________________________________________________
%
\begin{table}
\caption[]{A selection of lines in the $K$-band.}
 \label{tab:Klinelist}
 \begin{center}
{\changed
{\changedtwo
 \begin{tabular}{ll}  \hline \hline
 \noalign{\smallskip}
$\lambda_\mathrm{vac}$ [$\mu$m]  & Transition \\
  \noalign{\smallskip}
  \hline
  \noalign{\smallskip}
 %--------------------------------------------------
 2.0379      &   \ion{He}{ii} 15--8 \\
 2.0431      &   \ion{He}{i} 6p\,$^3$P--4s\,$^3$S \\
 2.0587      &   \ion{He}{i} 2p\,$^1$P--2s\,$^1$S \\
 2.0607      &   \ion{He}{i} 7d\,$^3$D--4p\,$^3$P \\
 2.0706/0802/0842& \ion{C}{iv} 3d\,$^3$D--3p\,$^3$P \\
 2.1081      &   \ion{C}{iii} 5p\,$^1$P--5s\,$^1$S \\
 2.1126      &   \ion{He}{i} 4s\,$^3$S--3p\,$^3$P \\
 2.1138      &   \ion{He}{i} 4s\,$^1$S--3p\,$^1$P \\
 2.14999     &   \ion{He}{i} 7s\,$^3$S--4p\,$^3$P \\
 2.1586      &   \ion{He}{i} 7p\,$^1$P--4d\,$^1$D \\
  \noalign{\smallskip}
\multicolumn{1}{l}{Br$\gamma$ blend:} \\
 2.1614      &   \ion{He}{i} 7f\,$^3$F--4d\,$^3$D \\
 2.1623      &   \ion{He}{i} 7f\,$^1$F--4d\,$^1$D \\
 2.1647      &   \ion{He}{i} 7g\,$^3$G--4f\,$^3$F \\
 2.1647      &   \ion{He}{i} 7g\,$^1$G--4f\,$^1$F \\
 2.1652      &   \ion{He}{ii} 14--8 \\
 2.1653      &   \ion{He}{i} 7d\,$^1$D--4f\,$^1$F \\
 2.1655      &   \ion{He}{i} 7d\,$^3$D--4f\,$^3$F \\
 2.1661      &   \ion{H}{i} 7--4 \\
  \noalign{\smallskip}
 2.1792      &   \ion{He}{ii} 23--9 \\
 2.1821      &   \ion{He}{i} 7p\,$^3$P--4d\,$^3$D \\
 2.1846      &   \ion{He}{i} 7d\,$^1$D--4p\,$^1$P \\
 2.1891      &   \ion{He}{ii} 10--7 \\
 2.2165      &   \ion{He}{ii} 22--9 \\
 2.2291      &   \ion{He}{i} 7s\,$^1$S--4p\,$^1$P \\
 2.2471/2513 &   \ion{N}{iii} 5p\,$^2$P--5s\,$^2$S \\
 2.2608      &   \ion{He}{ii} 21--9 \\
 2.30697     &   \ion{He}{i} 6p\,$^1$P--4s\,$^1$S \\
 2.3142      &   \ion{He}{ii} 20--9  \\
 2.3470      &   \ion{He}{ii} 13--8  \\
\hline %--------------------------------------------------
\end{tabular}
}}
\end{center}
\end{table}
%-------------------------------------------------------------
%
\subsection{PoWR -- WN model fitting}
\label{sec:fitting}
The normalized emission line spectra of Wolf-Rayet {\changed stars} depend 
predominantly on two parameters, the stellar temperature $T_\ast$ and 
a combination of parameters which was put in the form of the so-called
``transformed radius'' \citep{Schmutz+1989}
\begin{equation}
R_{\rm t} = R_* \left[ \frac{\varv_\infty}{2500\,\textrm{km~s}^{-1}}
  \left/ \frac{\sqrt{D}\dot{M}}{10^{-4}\,M_\odot\textrm{a}^{-1}}\right]
  \right. \,.
\label{eq:transformed-radius}
\end{equation}
The first step in the analysis is {\changed to} adjust these two
parameters such that 
the observed helium lines are reproduced. Concerning Br$\gamma$, the
hydrogen abundance is the relevant parameter, while the terminal
velocity $\varv_\infty$ sets the width of the line profiles.

Our analyses start from the two available grids\footnote{\tt
http://www.astro.physik.uni-potsdam.de/PoWR.html} of WN-type model
atmospheres, calculated for a hydrogen abundance of 20\% by mass
(``WNL'' grid) and for hydrogen-free stars (``WNE'' grid), respectively.
Since all our program stars are of late spectral type with signatures of
hydrogen in their spectra (WN9h), the WNL grid turns out to be more
appropriate. This grid had been calculated for a fixed terminal wind
velocity of $\varv_\infty = 1000$\,km~s$^{-1}$. After the best-fitting
grid models {\changed were} pointed out, we calculated individual
models for each star with specially adjusted $\varv_\infty$ and
hydrogen abundance in order to optimize the fits.

\begin{figure}[t]
   \centering
   \includegraphics[width=\columnwidth]{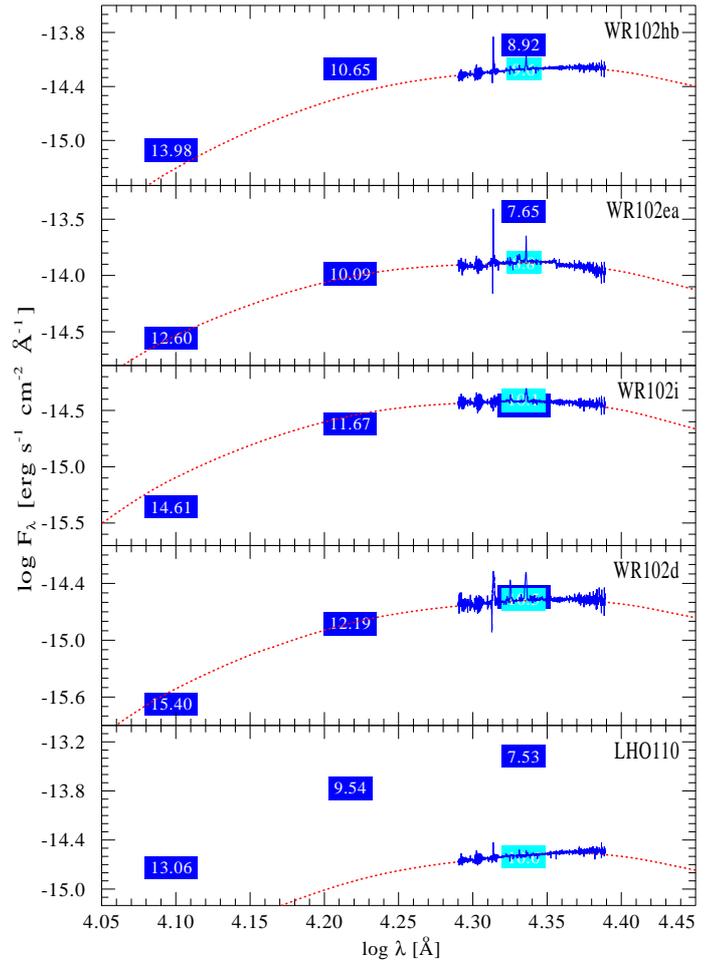}
   \caption{Spectral energy distributions for the sample stars. Blocks
     with labels indicate the 2MASS magnitudes for the 
   $J$-, $H$- and $K_\textrm s$-band (dark blue boxes), as well as the LHO
   $K_\textrm{s}$ magnitude (light blue box). The solid line is the 
   flux-calibrated LHO spectrum while the dotted line refers to the
   reddened model continuum.}
              \label{fig:SED}
   \end{figure}

\begin{figure*}
   \centering
   \includegraphics[width=\textwidth]{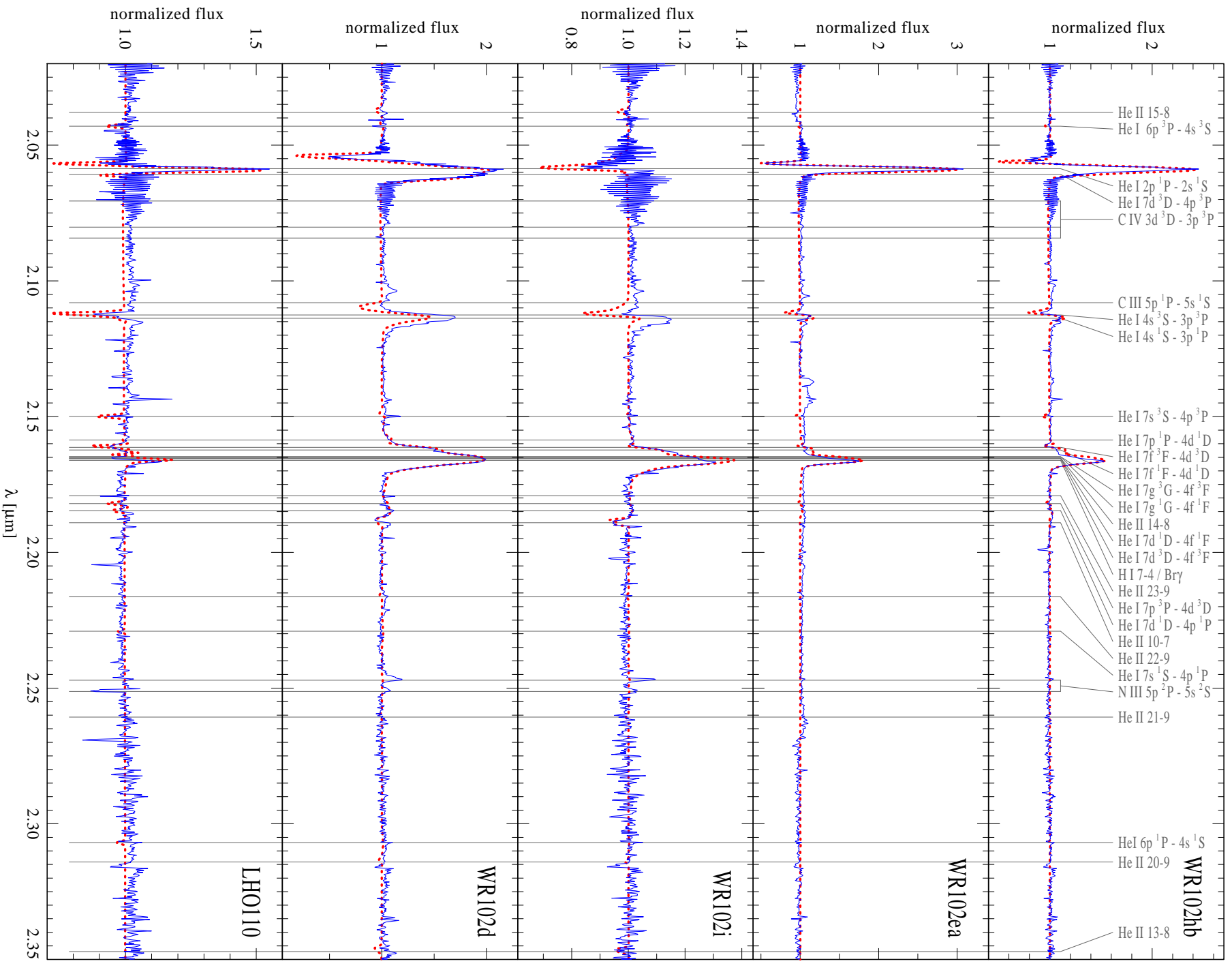}
   \caption{Spectral fits for the sample stars. Observed spectra
   (blue solid line) are normalized using their reddened model
   continuum, the normalized model spectrum is overlayed (red dotted
   lines). The observed spectra are corrected for radial velocities
   according to Table\,2 in the LHO catalog.}
              \label{fig:linespec}
   \end{figure*}

{\changed 
The results are shown in Table\,\ref{tab:WNparameters},} while the
observed spectra with the model fits are shown in
Fig.\,\ref{fig:SED} and
Fig.\,\ref{fig:linespec}. Note that we normalized the observed spectra
by division through the reddened model continuum.
%________________________________________________________________
%
\begin{table*}%[!h]
 \caption[]{Stellar parameters for the analyzed WN stars in the
   Quintuplet cluster}  
 \label{tab:WNparameters}
 \begin{center}
{\changed
 \begin{tabular}{rlcrrccrcrcccr}  \hline \hline
 \noalign{\smallskip}
  LHO & Alias & Spectral &$K_\textrm s$ & $T_*$& $\log R_\mathrm t$& $
  \varv_\infty$& $X_\mathrm H$& $E_{b- \varv}$& $M_K$& $R_*$& $\log
  \dot{M}$& $\log L$&$ \frac{\dot{M} \varv_\infty}{L/c}$ \\
  No. & & type & [mag] & [kK] &[$R_\odot$]& [km~s$^{-1}$]& [\%]& [mag]& [mag]&
  [$R_\odot$]& [$M_\odot$~a$^{-1}$]& [$L_\odot$]&  \\
  \noalign{\smallskip}
  \hline
  \noalign{\smallskip}
 67&WR\,102hb&WN9h& 9.6& 25.1& 1.55& 400& 19& 8.0& -8.3& 86& -4.52& 6.42& 0.2\\ 
 71&WR\,102ea&WN9h& 8.8& 25.1& 1.52& 300& 25& 6.1& -8.3& 83& -4.62& 6.39& 0.1\\ 
 99&WR\,102i &WN9h&10.1& 31.6& 1.65& 900& 45& 6.3& -7.1& 42& -4.79& 6.19& 0.5\\ 
158&WR\,102d &WN9h&10.5& 35.1& 1.11& 700& 15& 7.1& -7.0& 29& -4.32& 6.07& 1.4\\ 
110&         &WN9h&10.6& 25.1& 1.68& 300&  5& 8.7& -7.6& 67& -5.01& 6.20& 0.1\\ 
\hline
 \end{tabular}
}
 \end{center}
\end{table*}
%________________________________________________________________
%

%________________________________________________________________
%
%________________________________________________________________
%

%________________________________________________________________
%
%________________________________________________________________
%

In the next step, we fit the spectral energy distribution (SED) which
for our program stars is only available for the near-infrared range.
Figure\,\ref{fig:SED} shows the photometric fluxes in $J$, $H$ and $K$
from 2MASS \citep{2MASS}, and our flux-calibrated spectra from
{\changed the LHO catalog} together with the photometric magnitude
derived there. Crowding in the cluster field does not allow photometric
fluxes to be extracted, e.g. from {\em Spitzer} IRAC or MSX images.

In the case of \object{WR\,102d} we encountered a problem with the 2MASS
magnitudes from the final release of the catalog, which gives
unplausible values. Moreover, we find a coordinate offset between the
2MASS position and the LHO position of \object{WR\,102d} of almost
5\,\arcsec; we identify the 2MASS position rather with \object{LHO\,147} (see
LHO catalog). 
However, the data from the 2MASS {\em intermediate release} catalog, although
considered to be obsolete, contain a point source at exactly the
position of \object{WR\,102d} with fluxes that fit very well to our calibrated
$K$-band spectrum. Therefore we adopt these values for the SED
fit. {\changed The comparison with the near-IR photometry presented by
  \citet[filters $J$, $H\,2$ and $K\,2$]{FMM99} supports this choice.}

The model continuum (red dotted lines in Fig.\,\ref{fig:SED}) is
{\changed then} fitted to the observations by adjusting two
parameters, the reddening parameter (e.~g. {\changed given in the form
of the color excess $E_{b - \varv}$ in the narrow-band system of
\citet{Smith1968}, see Table\,\ref{tab:WNparameters}}) and the
stellar luminosity. {\changed We} adopt a distance to the \object{Quintuplet}
of 8\,kpc, corresponding to a distance modulus of 14.52\,mag
\citep{Reid1993}. The reddening law which we adopt is from
\citet{Moneti+2001}. {\changed The reddening parameter $E_{b - \varv}$
translates to the usual Johnson system as $E_{B - V} = 1.21 \times
E_{b - \varv}$, while the extinction in the $K$-band is $A_K = 0.42
\times E_{b - \varv}$.} Thanks to the approximate invariance of
Wolf-Rayet line spectra for models with same transformed radius,
models can be scaled to different luminosities for the SED fit while
the line fit is preserved. Generally we can reproduce the SED quite
well.  We attribute the remaining discrepancies to problems of the
2MASS photometry in the crowded field.

Remember that we have already employed the model continuum to normalize
the observed spectra. Insofar, the described two steps of the analysis
are in fact an iterative procedure. The derived extinction parameters
are compiled in Table\,\ref{tab:WNparameters} and further discussed in
Sect.\,\ref{sec:extinction}.
After we fitted the line spectrum and the SED, the whole set of 
stellar parameters is established (see Table\,\ref{tab:WNparameters}). 

{\changed 

\subsection{WR\,102ea and WR\,102hb}
These stars, as well as \object{LHO\,110} (see Sect.\,\ref{sec:LHO110}), do not
show clear \ion{He}{ii} lines in their spectra, which is indicative
for a low-temperature regime in their atmospheres. 
{\changed Attempts to fit} the spectra with models of too cool
temperatures makes those lines disappear completely and overestimates the
\ion{He}{i} lines instead, especially \ion{He}{i} 2.059\,$\mu$m.
In comparison, a hotter model produces too strong \ion{He}{ii} lines in
a strength in which they are not observed. 

Additionally, these stars have terminal wind velocities of
300-400\,km\,s$^{-1}$, which is rather slow for WR winds. In
comparison, the adopted standard microturbulence $\varv_{\rm D} =
100\,$km\,s$^{-1}$ {\changedtwo might be too high. 
Therefore, we tested different microturbulent velocities down to
$\varv_{\rm D} = 20\,$km\,s$^{-1}$. However, the best fit to the
observed profile shapes is obtained for $\varv_{\rm D} =
80$\,km\,s$^{-1}$. In any case, the effect of this parameter on the
synthetic $K$-band spectra is small.} 

\subsection{WR\,102d and WR\,102i}
The spectra of these two stars both show \ion{He}{i} and \ion{He}{ii}
lines which allow the temperature to be determined more reliably than
in the previous cases.  As for \object{WR\,102i}, however, our model fall short
in reproducing the strong emission feature of \ion{He}{i} at
2.113\,$\mu$m, while the absorption feature is too strong. Although a
bit hidden in the telluric residuals, the same seems to hold for
\ion{He}{i} at 2.059\,$\mu$m, but the \ion{He}{ii} absorption at
2.189\,$\mu$m fits perfectly. Alternatively, we found slightly hotter
models which reproduce the \ion{He}{i} emission, but then fail with
the \ion{He}{ii} lines turning into emission. We are not sure about
the reason for these inconsistencies, but we note that e.~g.\ the
\ion{N}{iii} doublet at 2.247/51\,$\mu$m seems to show a different
radial velocity, possibly indicating that the spectrum is composite
from a binary. This needs further observations to be confirmed. For
the moment, the fit problems introduce extra uncertainty which might
be lower for the alternative models.

In the case of \object{WR\,102d} the lines \ion{He}{i} at 2.059\,$\mu$m and
\ion{He}{ii} at 2.189\,$\mu$m can be reproduced consistently by the
model. With the stellar temperature being well constrained, the
Br$\gamma$ fit also yields an accurate hydrogen abundance (see
Fig.\,\ref{fig:H-content}). 

However, the spectra of both stars show remarkably strong \ion{N}{iii}
lines at 2.247/51\,$\mu$m.  Although we already spent some efforts in
improving the corresponding part in our nitrogen model atom, our
models notoriously fail to reproduces this doublet in the strength it
is observed here, while it is reasonably well fitted for the weak
emission in \object{WR\,102ea} and \object{WR\,102hb}.

In comparison to the other stars in the sample these two stars show
the highest terminal wind velocities, which are more representative
for WN stars. The final best fitting models were calculated with the
standard microturbulence $\varv_{\rm D} = 100\,$km\,s$^{-1}$.  }

\subsection{LHO\,110}
\label{sec:LHO110}
\object{LHO\,110} was classified as O6-8~Ife in the LHO catalog. The
emission line spectrum is very similar to the spectral appearance of the
WN stars, although the spectral features are less pronounced. 
In detail, the star has a similar $K_\textrm{s}$ magnitude as the WN9
stars and also shows similar line ratios for (\ion{He}{ii}
2.189\,$\mu$m)/(\ion{He}{ii}/Br$\gamma$ 2.166\,$\mu$m) and
(Br$\gamma$)/(\ion{He}{i} 2.059\,$\mu$m). 

We analyzed the star in the same way as described above for the WN
stars. We calculated tailored models to fit the line spectrum and SED
accordingly, see Fig.\,\ref{fig:SED} and \,\ref{fig:linespec}.
{\changed The obvious weakness of Br$\gamma$ leads to a hydrogen mass
fraction as low as 5\%. However, models with zero hydrogen cannot
fit the blend at all which excludes the star to be hydrogen free.
Hence the spectral classification in the LHO catalog must be revised
to WN9h. Following the naming conventions of \citet{vdH2001} this
star could become {\it WR\,102df}.}  The flux-calibrated spectrum
can be well fitted with the model SED, but cannot be connected to the
2MASS photometry. As in the case of \object{WR\,102d}, we attribute this to the
problems of the 2MASS measurements. The star lies in the central
region of the cluster where crowding might have affected the 2MASS
data.  Further on, a set of small absorption lines is present in the
spectrum, which we cannot identify.  They might indicate that
\object{LHO\,110} is a binary.  However, {\changed like the other
stars in our sample, \object{LHO\,110} has not been} detected
{\changed in X-rays} \citep{Wang+2006}. {\changed Collision of stellar
winds in} binary WN stars {\changed often make them brighter X-ray
sources than single stars \citep[e.~g.][]{Oskinova2005}. Therefore,
the binary nature of \object{LHO\,110} cannot yet be confirmed.}

{\changed
\subsection{Error margins}
The temperature of the model to be fitted can quite accurately be
determined for at least those sample stars that show \ion{He}{i} and
(even weak) \ion{He}{ii} lines.  In comparison, too cool or too hot a
model would produce \ion{He}{ii} lines in a weakness or strength in
which they are not observed. From this, we consider our temperatures
to be exact within the range of $\pm 0.025$\,dex, in general.

Different influences contribute to the error of the luminosities
given in Table\,\ref{tab:WNparameters}. The fit of the SED depends on
the accuracy of the photometric measurements. The 2MASS colors often
suffer from source confusion, while the $K_\textrm{s}$ magnitudes from the LHO
catalog have been calibrated to an accuracy of $\pm 0.1$\,mag.
The reddening parameter $E_{b - \varv}$ obtained from the slope of the SED is
accurate by about $\pm 0.1$\,mag, which gives only a 0.04mag uncertainty in
the $K$-band extinction.

Given that we fit the SED in the Rayleigh-Jeans domain, the bolometric
correction {\changedtwo factor ($10^{0.4 BC}$)} scales with
$T_*^3$. Hence the temperature contributes about 10\,\% to the error
margin of the luminosity. In combination we think that $\log{L}$ has a
typical error of about $\pm 0.16$\,dex.

%________________________________________________________________
%
   \begin{figure}[t]
   \centering
   \includegraphics[width=\columnwidth]{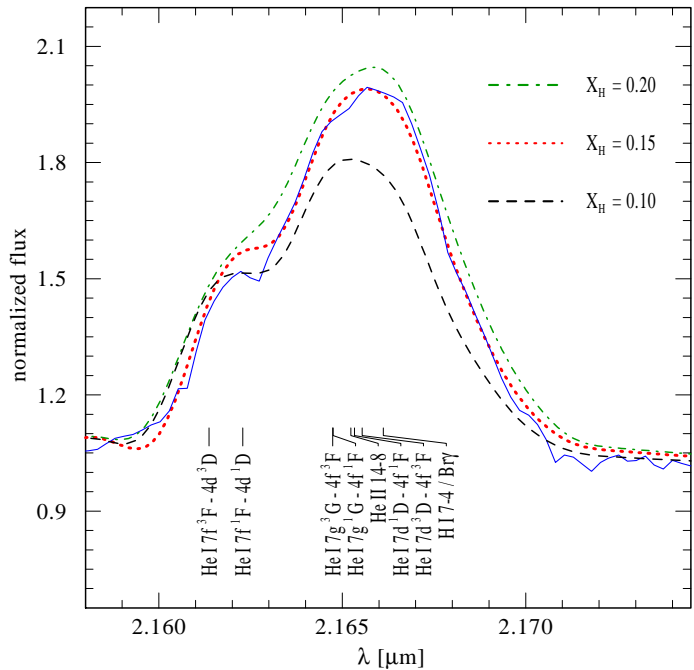}
   \caption{Normalized line spectrum of WR\,102d showing the
     {\changed Br$\gamma$\,/~\ion{He}{i}\,/~\ion{He}{ii}} blend (blue
     solid line) with the best fitting model (red dotted line). For
     comparison we plot models {\changed with 20\,\% (green
       dash-dotted line) and 10\,\% (black dashed line)} 
     hydrogen by mass but otherwise identical parameters.}
              \label{fig:H-content}
   \end{figure}
%________________________________________________________________
%

As described in Sect.\,\ref{sec:fitting}, the hydrogen content was
determined by fitting models with different mass fractions of hydrogen
to the spectra. Fig.\,\ref{fig:H-content} 
shows {\changed the Br$\gamma$\,/~\ion{He}{i}\,/~\ion{He}{ii}} blend
at 2.166\,$\mu$m of \object{WR\,102d}. The model with a hydrogen abundance of
{\changed 15\,\% by mass} gives the best fit to the observation (dotted
line) within the error range of 10 to 20\,\% hydrogen.
Therefore we conclude that the derived values for $X_\mathrm{H}$ are
correct within $\pm 5$\,\% by mass, accounting also for uncertainties
in the fit of the continuum.%} 
}

\subsection{Extinction}
\label{sec:extinction}
In the LHO catalog we assumed an average extinction of $A_K =
3.28$\,mag \citep{FMM99} for the \object{Quintuplet cluster}.  With the model
fit of the SEDs for our sample stars, we have determined individual
values for the reddening.  The results for the stars are listed in
Table\,\ref{tab:extinction}, with an average of $A_K = 3.1 \pm
0.5$\,mag, similar to the one determined by \citet{Barniske+2008}, who
analyzed the stars WR102c and WR102ka, and \citet{FMM99}.  The average
extinction of the \object{Arches cluster} was found to be $A_K = 2.8$\,mag by
\citet{Martins+2008}, slightly lower than for the \object{Quintuplet}.

%________________________________________________________________
%
\begin{table}[t]
  \caption[]{Extinction derived from fitting the model SEDs, applying
  the reddening law of \citet{Moneti+2001} in the infrared range.}
  \label{tab:extinction}
  \begin{center}
  \begin{tabular}{llll} \hline \hline
  Object & $A_V$ & $A_K$ & ref.\\
         & [mag] & [mag] \\
  \noalign{\smallskip} \hline
  \noalign{\smallskip}
   WR102hb, Q\,8 & {\changed 30}  & {\changed 3.4} & this work (t.w.) \\
   WR102ea, Q\,10& 23  & 2.6 & t.w. \\
   WR102i        & 24  & 2.7 & t.w. \\
   WR102d        & {\changed 27}  & 3.0 & t.w. \\
   LHO\,110      & 33  & 3.7 & t.w. \\
   mean          & 27 $\pm$ 4 & 3.1 $\pm$ 0.5 & t.w.\\ \hline 
  \noalign{\smallskip}
   WR102c        & 26 $\pm$ 1& 2.9 $\pm$ 0.1& \citet{Barniske+2008}\\
   WR102ka       & 27 $\pm$ 5& 3.0 $\pm$ 0.6 & \citet{Barniske+2008}\\ \hline
  \noalign{\smallskip}
   Quintuplet mean& 29 $\pm$ 5& 3.28 $\pm$ 0.5 & \citet{FMM99}\\
   Arches mean   & 24.9 & 2.8 & \citet{Martins+2008}\\ \hline
  \end{tabular}
  \end{center}
\end{table}
%________________________________________________________________
%

Within the errors, we conclude that the major part of the extinction
can be attributed to the foreground interstellar reddening. 
From the limited number of stars analyzed it
is not possible to derive an extinction map for the Quintuplet
cluster. We find a slight increase in the determined values towards
\object{LHO\,110} which is located closer in the cluster center.
This could be due to the intrinsic cluster extinction expected from
the presence of dust-producing WC stars \citep{Tuthill+2006}.

\section{Discussion}
\label{sec:discussion}

\subsection{Stellar parameters}
\begin{figure}
\centering
\includegraphics[width=\columnwidth]{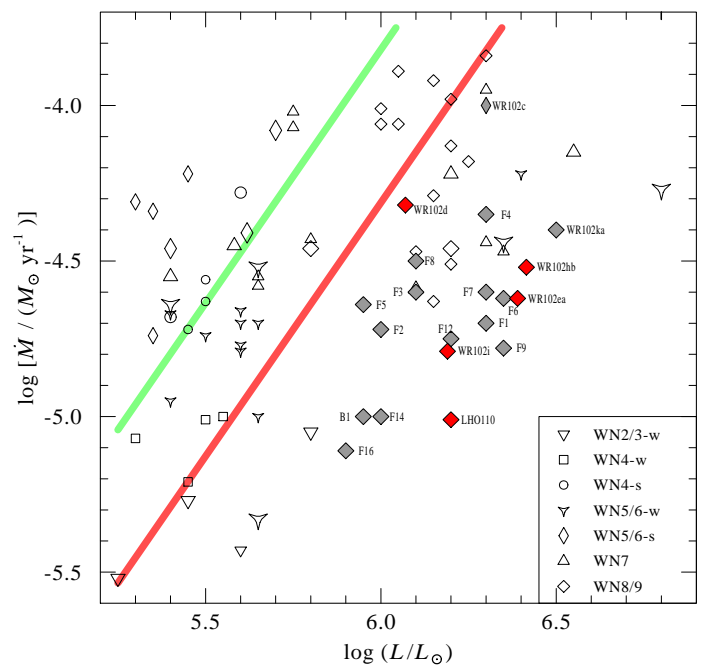}
\caption{Mass-loss rate versus stellar luminosity for the sample
   stars. For comparison we show Arches WNL stars from
   \citet[labeled with F and B numbers]{Martins+2008}, WR102c and
   WR102ka from \citet{Barniske+2008} and Galactic WN stars
   \citep[open symbols]{HGL2006}. The solid lines refer to the mass
   loss--luminosity relation of \citet{Nugis-Lamers2000} for
   hydrogen-free stars (green upper line) and stars containing 40\,\%
   hydrogen per mass (red lower line).} 
\label{fig:mdot}
\end{figure}

Fig.\,\ref{fig:mdot} shows the derived empirical mass-loss rates versus
stellar luminosity. The stars analyzed in this work fall below the mass
loss--luminosity relation of \citet{Nugis-Lamers2000} for stars
containing 40\,\% hydrogen per mass (lower line). From their determined
hydrogen abundance (see Table\,\ref{tab:WNparameters}), one would expect to find
these stars positioned between this relation and the one for
hydrogen-free stars (upper line). 

%________________________________________________________________
%________________________________________________________________

%________________________________________________________________
\begin{figure}[!ht]
\centering
\includegraphics[width=\columnwidth]{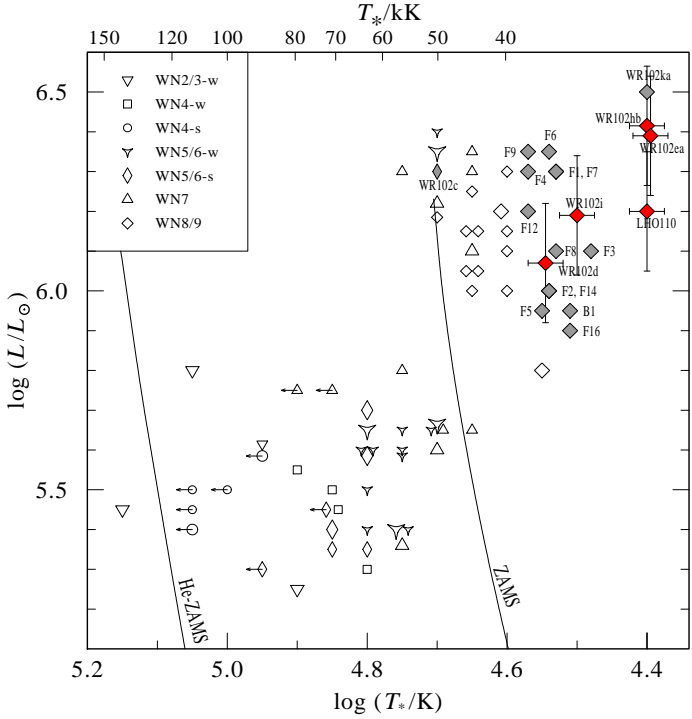}
\caption{HRD showing the analyzed WN9h stars (red filled
   symbols). For comparison we plot {\changed single} Galactic WN stars from
   \citet{HGL2006} (open symbols), WNL stars of the Arches cluster from
   \citet[labeled with F and B numbers]{Martins+2008}, and WR102c and
   WR102ka from \citet{Barniske+2008}. Different symbols refer to
   different spectral subtypes, cf. inset. Hydrogen and helium
   zero-age main-sequence are given as solid lines.}
\label{fig:hrd}
\end{figure}
%________________________________________________________________
%

Not only our program stars have rather low mass-loss rates compared to
their luminosities, the same holds for other Galactic WN stars shown
in Fig.\,\ref{fig:mdot} \citep[open symbols, from][]{HGL2006},
including the WNL stars in the \object{Arches cluster} \citep[][symbols labeled
B and F, from]{Martins+2008} which have similar hydrogen contents. For
radiation-driven winds, the mass-loss rates scale with the metallicity
\citep{GH2008}. Therefore, the relatively low mass-loss rates provide
an indirect argument that the metallicity in our program stars is not
significantly higher than in other Galactic WN stars studied so far.

{\changed However, the possibility that the Galactic center might be
metal enriched is still under debate. \citet{Najarro+2004}
obtained solar metallicities for a sample of stars in the 
\object{Arches cluster}, while recent results on a larger sample in
the same cluster favored slight metal enrichment in the range of 1.3
to 1.4\,$Z_\odot$ \citep{Martins+2008}.
For the \object{Quintuplet cluster}, \citet{Najarro+2009}
used quantitative spectral analysis of the LBV Pistol star and LBV
candidate \object{FMM\,362} to measure the
metallicity in the cluster. Their results indicate solar iron abundance and
roughly twice the solar abundance in the
$\alpha$-elements. 
They suggest that the enrichment in $\alpha$-elements versus Fe found
in the two Quintuplet LBVs is consistent with a top-heavy initial mass
function (IMF) in the GC.
\citet{Davies+2009} analyzed the red supergiant \object{GMM\,7}
(\object{LHO\,7}) in the cluster. The authors find a slight iron
enrichment which they explain by the evolved nature of the object and
the resulting hydrogen depletion. They argue that the ``initial'' iron
abundance would be in agreement with solar values.}

{\changed The Hertzsprung-Russell diagram (HRD, Fig.\,\ref{fig:hrd}) shows the
program stars of the \object{Quintuplet cluster} in comparison to other
Galactic WN stars. We include WN stars from the \object{Arches cluster}
\citep[][symbols labels B and F, from]{Martins+2008} and the two
bright stars WR102c and WR102ka \citep{Barniske+2008}. The whole
sample of single Galactic WN stars \citep[open symbols]{HGL2006} is
also shown for comparison. 

Obviously, the GC stars populate a special domain in the HRD.  They
lie within a group of rather cool, $T_* \le 50\,000$\,K, but very
luminous WN stars with $\log{(L/L_\odot)} > 6.0$. Like most WNL stars
in the Galaxy, our GC stars show some rest of hydrogen with mass
fractions of the order of 20\%. } {\changed While about half of the
Galactic WN stars belong to the ``early'' subclass (WNE) that resides
to the left side of the ZAMS (see Fig.\,\ref{fig:hrd}), no single such
hot WN star has been found in the Galactic center region.

One may wonder if the low stellar temperatures obtained for the GC stars
are due to systematic errors of spectral analyses that are based on
the $K$-band only. But the stars in the \object{Arches cluster} analyzed by
\citet{Martins+2008} show that independent groups arrive at
similar results. 

Can the lack of WNE stars be attributed to selection effects? Compared
to WNL stars, WNE stars are roughly three magnitudes fainter in the
$K$-band \citep[cf.][the difference is similar as for the visual
band]{HGL2006}. Since the apparent $K$-band magnitude of our program
stars is about 10\,mag (Table\,\ref{tab:WNparameters}), the WNE stars
should not all have escaped from detection in the LHO catalog which is
complete to the 13$^{\rm th}$ magnitude.

{\changedtwo
A statistical comparison with the WNE star population in other regions
of the Milky Way confirms that WNEs are usually detected with slightly
fainter $K$-band magnitudes compared to their local WNL
counterparts. However, the number ratio
$N_\mathrm{WNL}:N_\mathrm{WNE}$ varies largely from 24:33 for Galactic
WN \citep{HGL2006}, 4:11 for Wd~1 \citep{Crowther+2006}, 10:3 for the GC
Central cluster \citep{Martins+2007}, 13:0 for the \object{Arches}
\citep{Martins+2008}, and 5:0 for the \object{Quintuplet}, indicating a lack of
WNE stars in the GC region. 
}

Hence, we must conclude that the population of WN stars in the
Galactic center region is different from the Galactic WN population in
general. This may be attributed to various reasons: a special
star-formation history, a different initial mass function (IMF), or a
(slightly) higher metallicity that increases the mass loss during
stellar evolution. A closer investigation of this question may give
important insight into the conditions of the Galactic center
environment. }

For all stars in the sample we derive the wind momentum efficiency
number $\eta = \dot{M}\varv_\infty c/L$, see
Table\,\ref{tab:WNparameters}.  {\changedtwo Recall that we have
adopted a clumping contrast of $D=4$ throughout this paper (cf.\
Sect.\,\ref{sec:models}). If $D$ is higher, this would reduce the
empirical mass-loss rates $\propto D^{-1/2}$ and thus lead to
smaller $\eta$ values. However, this} number cannot exceed unity in
radiation-driven winds unless multiple scattering is taken
account. \citet{GH2005, GH2008} have shown that $\eta$-values of a few
can be easily obtained in hydrodynamically consistent WR models with
full radiative transfer. For all stars we find values of $\eta$ close
to unity, which is consistent with these stellar winds being
radiatively driven.

\subsection{Radio mass-loss rates}
\label{sec:radio}
Three of the sample stars are known to be radio sources from
\citet{Lang+2005}. We calculated the mass-loss rates from radio
free-free emission with radio fluxes from these authors and follow
\citet{Wright+Barlow1975}
\begin{equation}
\dot M =  \varv_{\infty}\,\frac{f_{\nu}^{3/4}\,d^{3/2}}{(23.2)^{3/4}\,
                (\gamma\, g_\mathrm{ff}\, \nu)^{1/2}} \left (
                \frac{\mu}{Z} \right )\,.
\end{equation}
The free-free Gaunt factor $g_\mathrm{ff}$ is derived from the relation
by \citet{Leitherer+Robert1991} with an assumed electron temperature
of 10\,000\,K. Terminal velocity $\varv_\infty$ and mean molecular
weight $\mu$ are taken from our analysis (see Table\,\ref{tab:radio}).
Assuming that helium stays ionized in the radio emitting region of the
stellar wind, the mean number of electrons is set to $\gamma =1$. 
Radio mass-loss rates were calculated from the radio flux at 22.5\,GHz
for QR\,5 (WR102ea) and QR\,8 (WR102d), and at 8.5\,GHz for QR\,4
(\object{LHO\,110}). This allows the derived mass-loss rates to be compared
directly with those obtained by \citet{Lang+2005}, but 
note that their original values for $\dot{M}$, given in parenthesis in
Table\,\ref{tab:radio}, were calculated with different assumptions on
the terminal velocity and the mean molecular weight. 

%________________________________________________________________
\begin{table}[t]
  \caption[]{Mass-loss rates derived from radio
  free-free emission ($\dot{M}_\mathrm{radio}$), compared to those
  determined from the $K$-band spectra in the present paper
  ($\dot{M}_\mathrm{IR}$). Based on radio measurements from
  \citet{Lang+2005}, we derive $\dot{M}_\mathrm{radio}$ consistently
  with the parameters determined in this paper. The values in
  parenthesis are the mass-loss rates from \citet{Lang+2005}
  who adopted $\varv_\infty=1000$\,km~s$^{-1}$ and $\mu =2$ for all stars.}
  \label{tab:radio}
  \begin{center}
{\changed
  \begin{tabular}{llllll} \hline \hline
  \noalign{\smallskip}
   Object  & alias & $\varv_\infty$& $\mu$ & $\log{\dot{M}_\mathrm{radio}}$ &
           $\log{\dot{M}_\mathrm{IR}}$\\ 
           &       & [km~s$^{-1}$]&&[$M_\odot$~a$^{-1}$]&[$M_\odot$~a$^{-1}$]\\
  \noalign{\smallskip} \hline
  \noalign{\smallskip}
   WR102ea & QR\,5 & 300.& 2.30& -4.44~(-3.96) & -4.62 \\
   WR102d  & QR\,8 & 700.& 2.78& -4.30~(-4.37) & -4.32 \\ 
   LHO\,110& QR\,4 & 300.& 3.52& -4.12~(-3.82) & -5.01 \\ \hline
  \end{tabular}
}
  \end{center}
\end{table}
%________________________________________________________________
%

For QR\,5 (WR102ea) and QR\,8 (WR102d) the mass-loss rates which we
deduce from the radio measurements by \citet{Lang+2005} are in
excellent agreement with those obtained in the present paper from the
$K$-band analysis (see Table\,\ref{tab:radio}).  Thus, these two radio
detections can be confirmed as stellar wind sources.

It has to be noted that the radio mass-loss rates quoted in
Table\,\ref{tab:radio} are derived for an unclumped medium (clumping
contrast $D=1$), while the model $K$-band spectra are calculated for
slightly clumped winds ($D = 4$).  {\changedtwo Hence, the good
agreement implies that the degree of clumping generally decreases
from the line forming region towards the radio emitting as
previously found for OB stars \citep{Puls+2006} and WR stars
\citep{Nugis+1998, Liermann-Hamann2008}.}

In the case of \object{LHO\,110}, however, the measured radio flux is about ten
times higher then expected from the \citet{Wright+Barlow1975} model
compared to the mass-loss rate derived from the $K$-band analysis.
This might indicate that the radio emission is produced in a
colliding-wind zone. On the other hand, binaries often show a
non-thermal radio spectrum with a negative radio spectral index. For
\object{LHO\,110}, \citet{Lang+2005} found a spectral index of +1.4, which is
closer to the prediction of the \citet{Wright+Barlow1975} theory for
thermal wind emission (spectral index +0.6).  Considering these
contradictory indications, the nature of the radio emission and the
binary status of \object{LHO\,110} remains unsettled.

\subsection{LBV candidates?}
{\changed We find two stars in our sample, WR102ea and WR102hb} with
very high luminosities and low terminal velocities.  Their mass-loss
rates are comparable to those derived for the Pistol star and \object{FMM\,362}
\citep{Najarro+2009}, or WR102ka \citep{Barniske+2008} thus leading to
the conclusion the stars might be LBV candidate stars.  For \object{LHO\,110}
the {\changed slightly smaller luminosity} sets the star in the HRD
just {\changed below} the Humphreys-Davidson limit adapted from
\citet[their Fig\,12]{Figer+1998b}, while other Quintuplet LBV
candidates are above this empirical relation (see
Fig.\,\ref{fig:lbv}). Also the very low mass-loss rate might question
the LBV scenario for {\changed \object{LHO\,110}}, despite its spectral
appearance.

\begin{figure}[t]
   \centering
   \includegraphics[width=\columnwidth]{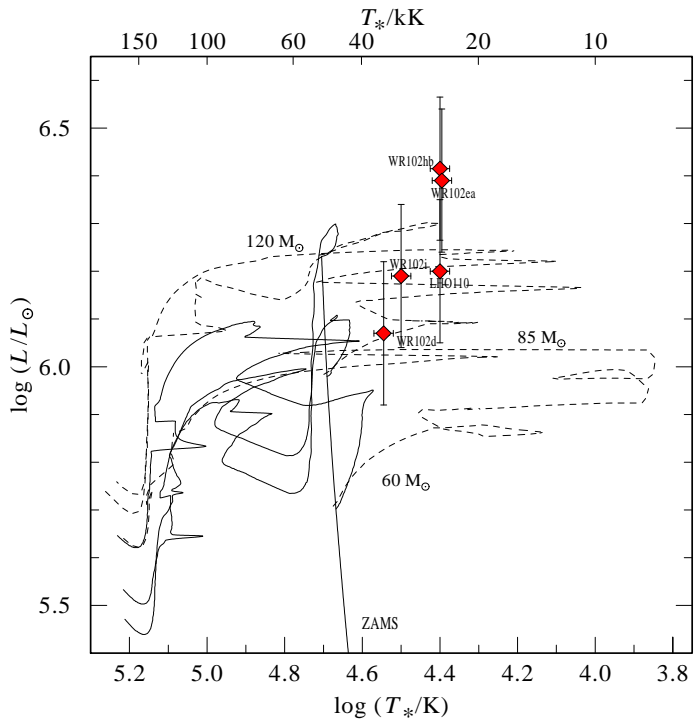}
   \caption{HRD with evolutionary tracks from
   \citet{Meynet+Maeder2003} for $M_{\rm ini}$ of 60, 85 and
   120\,$M_\odot$. The dashed lines refer to tracks without
   rotation, solid lines to tracks with $\varv_{\rm ini} =
   300$\,km~s$^{-1}$. Symbols are for the analyzed Quintuplet WNL
   stars as in Fig.\,\ref{fig:hrd}.}
              \label{fig:hrdwithtracks}
   \end{figure}

\citet{Clark+2005LBV} proposed an empirical limit for LBV (candidate)
stars in the quiescent state. This hot LBV minimum light strip is
defined up to luminosities of $\log(L/L_\odot) \sim 6.2$ by observed
LBVs and LBV candidates, but a possible extension towards higher
luminosities is discussed. {\changed Further on, \citet{Clark+2005LBV}
argue} that LBV stars in this part of the HRD will look like WNL
stars. {\changed As recently presented by \citet{Groh+2009b}, this
minimum light strip is connected to the strong-variable LBVs
suspected to be fast rotators reaching their critical rotational
velocity, e.~g. the known strong variable LBVs AG~Car and HR~Car lie
at this limit. In contrast slow-rotating LBV stars are assumed to be
less variable.  In any case, we don't find hints in the spectra of
our sample stars to be fast rotating and the question of variability
cannot be settled without further observations.  Still, all three
stars, \object{WR\,102ea}, \object{WR\,102hb}, and \object{LHO\,110},
are thus left in the possible LBV regime within the HRD.

%________________________________________________________________
%
   \begin{figure}[t]
   \centering
   \includegraphics[width=\columnwidth]{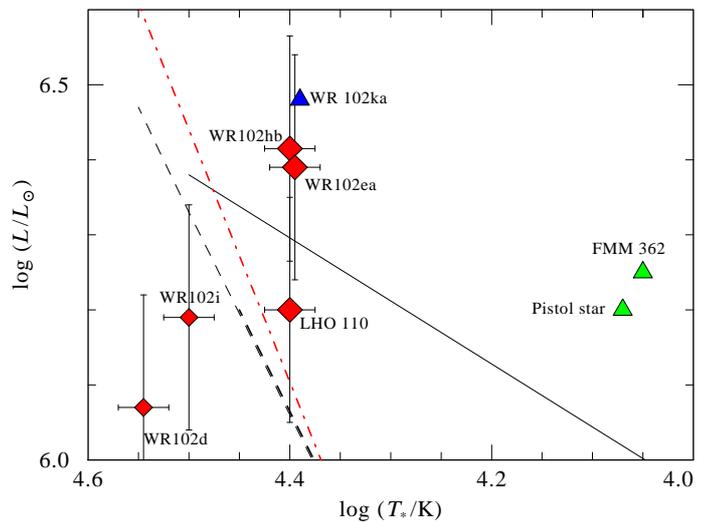}
   \caption{HRD with LBV (candidate) stars in the vicinity of the Quintuplet
   cluster. {\changed In addition to the program stars of this paper
     we indicate the Pistol star and FMM\,362 \citep[parameters taken
     from][]{Najarro+2009} and WR\,102ka \citep{Barniske+2008}.
    The Humphreys-Davidson limit
   \citep[solid line, adapted from][]{Figer+1998b} is shown as well as
   the hot LBV minimum light strip from \citet[dashed,
   extrapolated with thin dashes to higher
   luminosities]{Clark+2005LBV}. Even the steeper version of this
     latter limit from \citet[dash-dotted line]{Groh+2009b} puts
     WR\,102ea, WR\,102hb, LHO\,110, and WR102ka to its right side, i.e. in the
     suggested LBV domain.} } 
              \label{fig:lbv}
   \end{figure}
%________________________________________________________________
%

WR102ea and WR102hb both are {\changed placed} above the
evolutionary track with 120\,$M_\odot$ initial mass, see
Sect.\,\ref{sec:evolution} and Fig.\,\ref{fig:hrdwithtracks}. But as 
described recently by \citet{Maeder+2008}, such massive stars might
skip the LBV phase in their evolution completely.
}

%
%________________________________________________________________
%________________________________________________________________
%

\subsection{Stellar evolution}
\label{sec:evolution}
The HRD with evolutionary tracks from \citet{Meynet+Maeder2003} is
shown in Fig.\,\ref{fig:hrdwithtracks}. Most sample stars lie within
the tracks for solar metallicity between 60 and 120\,$M_\odot$ initial
mass implying that the WN stars are descendants from the most massive
O stars.  The tracks without rotation (dashed lines) extend further
into the cooler temperature domain than the tracks with rotation
(solid lines).  This might imply that the initial equatorial
rotational velocity of the sample stars was smaller than the assumed
300\,km~s$^{-1}$ in the evolutionary models.  Similar results were
already encountered when comparing the Galactic WN stars with the
Geneva evolutionary tracks \citep{HGL2006}. {\changed This indicates
that the stellar evolution of massive stars is not yet understood
fully.}  Two stars in the sample, \object{WR\,102hb} and
\object{WR\,102ea}, lie above the most massive track with
120\,$M_\odot$ initial mass.  We compare 
the stellar parameters of the stars with evolutionary tracks by
\citet{Langer+1994} as shown by \citet[Fig.\,15a]{Figer+1998b}. Thus
the initial stellar mass for these stars would be in the range $150
\le M_\mathrm{init} \le 200\,M_\odot$.

For those stars that can be {\changed assigned} to Geneva evolutionary
tracks, we estimated the age of the stars to be 2.1 to 3.6\,Ma and
find present-day masses of 30 to 45\,$M_\odot$. This is in agreement
with masses we derive following the relation by \citet{Langer1989},
which is strictly for pure helium stars only.  Considering the
determined stellar temperatures and luminosities we find {\changed
from the Geneva models} that the stars are still hydrogen burning
objects, in spite of their small hydrogen abundances.

\section{Summary}
\label{sec:conclusion}
We present the analysis of {\changed five} WN9h stars in
the \object{Quintuplet cluster}. $K$-spectra were taken from the LHO catalog
and fitted with tailored Potsdam Wolf-Rayet models for expanding
atmospheres (PoWR) to derive the fundamental stellar parameters.  
All studied stars are very luminous ($\log{(L/L_\odot)} > 6.0$), while
their stellar temperature corresponds to their late WN subtype
{\changed ($ T_* \le 35$\,kK). Along with other WNL stars in the Milky
Way the \object{Quintuplet} stars form a distinct group that occupies a
specific region in the HRD, separate from the WNE stars.} They still
contain {\changed a significant amount of hydrogen (up to 45\,\% per
mass)} in their atmospheres and show {\changed typical WN-like}
mass-loss rates.  

Radio detections exist for three of our sample stars. In two cases,
\object{WR\,102d} (\object{QR\,8}) and \object{WR\,102ea}
(\object{QR\,5}), the radio emission agrees with 
the prediction for free-free emission from their stellar winds. In
contrast, \object{LHO\,110} (\object{QR\,4}) is much brighter in the radio than
consistent with these assumptions, which possibly indicates that it
might be a binary with colliding winds, {\changed although none of the
sample stars has been detected in X-rays}.

The two most luminous stars of our sample correspond to initial masses
above 150\,$M_\odot$. The other three, less luminous stars can be
compared with detailed evolutionary tracks, revealing initial masses
{\changed in the range} of 60 to 120\,$M_\odot$, and ages of about
2.1-3.6 million years. As in the case of other Galactic WN stars, the
evolutionary tracks without rotation seem to fit better to the
{\changed stars' location in the} empirical HRD. According to the
evolutionary models, these stars are still core-hydrogen burning
objects.

From the analysis of the individual stars we obtained the interstellar
reddening and extinction. Based on the small sample, no extinction map
could be established but an average extinction of $A_K = 3.1 \pm 0.5$\,mag
($A_V = 27 \pm 4$\,mag) was determined towards the \object{Quintuplet cluster}.

\begin{acknowledgements}
  {\changed We are very thankful to our referee F. Najarro for his
    useful comments which helped to improve this manuscript.}
  This publication makes use of data products from the Two Micron All
  Sky Survey, which is a joint project of the University of
  Massachusetts and the Infrared Processing and Analysis
  Center/California Institute of Technology, funded by the National
  Aeronautics and Space Administration and the National Science
  Foundation. 
  This research has made use of the SIMBAD database, operated at CDS,
  Strasbourg, France.  
  L.~M. Oskinova is supported by the Deutsches Zentrum f\"ur Luft- und
  Raumfahrt (DLR) under grant 50~OR~0804.
  A. Liermann is supported by the Deutsche Forschungsgemeinschaft (DFG)
  under grant HA~1455/19.
\end{acknowledgements}

%\bibliographystyle{aa}
%\bibliography{/homes/liermann/literature/library}

\begin{thebibliography}{52}
\expandafter\ifx\csname natexlab\endcsname\relax\def\natexlab#1{#1}\fi

\bibitem[{{Barniske} {et~al.}(2008){Barniske}, {Oskinova}, \&
  {Hamann}}]{Barniske+2008}
{Barniske}, A., {Oskinova}, L.~M., \& {Hamann}, W.-R. 2008, \aap, 486, 971

\bibitem[{{Blum} {et~al.}(2001){Blum}, {Schaerer}, {Pasquali},
  {Heydari-Malayeri}, {Conti}, \& {Schmutz}}]{Blum+2001}
{Blum}, R.~D., {Schaerer}, D., {Pasquali}, A., {et~al.} 2001, \aj, 122, 1875

\bibitem[{{Clark} {et~al.}(2005){Clark}, {Larionov}, \&
  {Arkharov}}]{Clark+2005LBV}
{Clark}, J.~S., {Larionov}, V.~M., \& {Arkharov}, A. 2005, \aap, 435, 239

\bibitem[{{Crowther}(2007)}]{Crowther2007}
{Crowther}, P.~A. 2007, \araa, 45, 177

\bibitem[{{Crowther} {et~al.}(2006){Crowther}, {Hadfield}, {Clark},
  {Negueruela}, \& {Vacca}}]{Crowther+2006}
{Crowther}, P.~A., {Hadfield}, L.~J., {Clark}, J.~S., {Negueruela}, I., \&
  {Vacca}, W.~D. 2006, \mnras, 372, 1407

\bibitem[{{Crowther} {et~al.}(1995){Crowther}, {Smith}, {Hillier}, \&
  {Schmutz}}]{Crowther-etal1995}
{Crowther}, P.~A., {Smith}, L.~J., {Hillier}, D.~J., \& {Schmutz}, W. 1995,
  A\&A, 293, 427

\bibitem[{{Davies} {et~al.}(2009){Davies}, {Origlia}, {Kudritzki}, {Figer},
  {Rich}, \& {Najarro}}]{Davies+2009}
{Davies}, B., {Origlia}, L., {Kudritzki}, R.-P., {et~al.} 2009, \apj, 694, 46

\bibitem[{{Eckart} {et~al.}(2004){Eckart}, {Moultaka}, {Viehmann},
  {Straubmeier}, \& {Mouawad}}]{Eckart+2004}
{Eckart}, A., {Moultaka}, J., {Viehmann}, T., {Straubmeier}, C., \& {Mouawad},
  N. 2004, \apj, 602, 760

\bibitem[{{Figer} {et~al.}(1999{\natexlab{a}}){Figer}, {McLean}, \&
  {Morris}}]{FMM99}
{Figer}, D.~F., {McLean}, I.~S., \& {Morris}, M. 1999{\natexlab{a}}, \apj, 514,
  202

\bibitem[{{Figer} {et~al.}(1999{\natexlab{b}}){Figer}, {Morris}, {Geballe},
  {Rich}, {Serabyn}, {McLean}, {Puetter}, \& {Yahil}}]{FMG99}
{Figer}, D.~F., {Morris}, M., {Geballe}, T.~R., {et~al.} 1999{\natexlab{b}},
  \apj, 525, 759

\bibitem[{{Figer} {et~al.}(2002){Figer}, {Najarro}, {Gilmore}, {Morris}, {Kim},
  {Serabyn}, {McLean}, {Gilbert}, {Graham}, {Larkin}, {Levenson}, \&
  {Teplitz}}]{Figer+2002}
{Figer}, D.~F., {Najarro}, F., {Gilmore}, D., {et~al.} 2002, \apj, 581, 258

\bibitem[{{Figer} {et~al.}(1998){Figer}, {Najarro}, {Morris}, {McLean},
  {Geballe}, {Ghez}, \& {Langer}}]{Figer+1998b}
{Figer}, D.~F., {Najarro}, F., {Morris}, M., {et~al.} 1998, \apj, 506, 384

\bibitem[{{Geballe} {et~al.}(2000){Geballe}, {Najarro}, \&
  {Figer}}]{Geballe+2000}
{Geballe}, T.~R., {Najarro}, F., \& {Figer}, D.~F. 2000, \apjl, 530, L97

\bibitem[{{Gr{\"a}fener} \& {Hamann}(2005)}]{GH2005}
{Gr{\"a}fener}, G. \& {Hamann}, W.-R. 2005, \aap, 432, 633

\bibitem[{{Gr{\"a}fener} \& {Hamann}(2008)}]{GH2008}
{Gr{\"a}fener}, G. \& {Hamann}, W.-R. 2008, \aap, 482, 945

\bibitem[{{Gr{\"a}fener} {et~al.}(2002){Gr{\"a}fener}, {Koesterke}, \&
  {Hamann}}]{Graefener+2002}
{Gr{\"a}fener}, G., {Koesterke}, L., \& {Hamann}, W.-R. 2002, \aap, 387, 244

\bibitem[{{Groh} {et~al.}(2009){Groh}, {Damineli}, {Hillier}, {Barb{\'a}},
  {Fern{\'a}ndez-Laj{\'u}s}, {Gamen}, {Mois{\'e}s}, {Solivella}, \&
  {Teodoro}}]{Groh+2009b}
{Groh}, J.~H., {Damineli}, A., {Hillier}, D.~J., {et~al.} 2009, \apjl, 705, L25

\bibitem[{{Hamann} {et~al.}(1991){Hamann}, {Duennebeil}, {Koesterke},
  {Wessolowski}, \& {Schmutz}}]{Hamann+1991}
{Hamann}, W., {Duennebeil}, G., {Koesterke}, L., {Wessolowski}, U., \&
  {Schmutz}, W. 1991, \aap, 249, 443

\bibitem[{{Hamann} \& {Gr{\" a}fener}(2004)}]{Hamann-Graefener2004}
{Hamann}, W.-R. \& {Gr{\" a}fener}, G. 2004, A\&A, 427, 697

\bibitem[{{Hamann} {et~al.}(2006){Hamann}, {Gr{\"a}fener}, \&
  {Liermann}}]{HGL2006}
{Hamann}, W.-R., {Gr{\"a}fener}, G., \& {Liermann}, A. 2006, \aap, 457, 1015

\bibitem[{{Hillier} \& {Miller}(1998)}]{Hillier-Miller1989}
{Hillier}, D.~J. \& {Miller}, D.~L. 1998, \apj, 496, 407

\bibitem[{{Krabbe} {et~al.}(1991){Krabbe}, {Genzel}, {Drapatz}, \&
  {Rotaciuc}}]{Krabbe+1991}
{Krabbe}, A., {Genzel}, R., {Drapatz}, S., \& {Rotaciuc}, V. 1991, \apjl, 382,
  L19

\bibitem[{{Krabbe} {et~al.}(1995){Krabbe}, {Genzel}, {Eckart}, {Najarro},
  {Lutz}, {Cameron}, {Kroker}, {Tacconi-Garman}, {Thatte}, {Weitzel},
  {Drapatz}, {Geballe}, {Sternberg}, \& {Kudritzki}}]{Krabbe+1995}
{Krabbe}, A., {Genzel}, R., {Eckart}, A., {et~al.} 1995, \apjl, 447, L95

\bibitem[{{Lang} {et~al.}(2005){Lang}, {Johnson}, {Goss}, \&
  {Rodr{\'{\i}}guez}}]{Lang+2005}
{Lang}, C.~C., {Johnson}, K.~E., {Goss}, W.~M., \& {Rodr{\'{\i}}guez}, L.~F.
  2005, \aj, 130, 2185

\bibitem[{{Langer}(1989)}]{Langer1989}
{Langer}, N. 1989, A\&A, 210, 93

\bibitem[{{Langer} {et~al.}(1994){Langer}, {Hamann}, {Lennon}, {Najarro},
  {Pauldrach}, \& {Puls}}]{Langer+1994}
{Langer}, N., {Hamann}, W.-R., {Lennon}, M., {et~al.} 1994, \aap, 290, 819

\bibitem[{{Leitherer} \& {Robert}(1991)}]{Leitherer+Robert1991}
{Leitherer}, C. \& {Robert}, C. 1991, \apj, 377, 629

\bibitem[{{Liermann} \& {Hamann}(2008)}]{Liermann-Hamann2008}
{Liermann}, A. \& {Hamann}, W.-R. 2008, in Clumping in Hot-Star Winds, ed.
  W.-R. {Hamann}, A.~{Feldmeier}, \& L.~M. {Oskinova}, 247

\bibitem[{{Liermann} {et~al.}(2009){Liermann}, {Hamann}, \&
  {Oskinova}}]{Liermann+2009}
{Liermann}, A., {Hamann}, W.-R., \& {Oskinova}, L.~M. 2009, \aap, 494, 1137

\bibitem[{{Maeder} {et~al.}(2008){Maeder}, {Meynet}, {Ekstr{\"o}m}, {Hirschi},
  \& {Georgy}}]{Maeder+2008}
{Maeder}, A., {Meynet}, G., {Ekstr{\"o}m}, S., {Hirschi}, R., \& {Georgy}, C.
  2008, in IAU Symposium, Vol. 250, IAU Symposium, 3--16

\bibitem[{{Martins} {et~al.}(2007){Martins}, {Genzel}, {Hillier}, {Eisenhauer},
  {Paumard}, {Gillessen}, {Ott}, \& {Trippe}}]{Martins+2007}
{Martins}, F., {Genzel}, R., {Hillier}, D.~J., {et~al.} 2007, \aap, 468, 233

\bibitem[{{Martins} {et~al.}(2008){Martins}, {Hillier}, {Paumard},
  {Eisenhauer}, {Ott}, \& {Genzel}}]{Martins+2008}
{Martins}, F., {Hillier}, D.~J., {Paumard}, T., {et~al.} 2008, \aap, 478, 219

\bibitem[{{Meynet} \& {Maeder}(2003)}]{Meynet+Maeder2003}
{Meynet}, G. \& {Maeder}, A. 2003, \aap, 404, 975

\bibitem[{{Moneti} {et~al.}(2001){Moneti}, {Stolovy}, {Blommaert}, {Figer}, \&
  {Najarro}}]{Moneti+2001}
{Moneti}, A., {Stolovy}, S., {Blommaert}, J.~A.~D.~L., {Figer}, D.~F., \&
  {Najarro}, F. 2001, \aap, 366, 106

\bibitem[{{Najarro} {et~al.}(2009){Najarro}, {Figer}, {Hillier}, {Geballe}, \&
  {Kudritzki}}]{Najarro+2009}
{Najarro}, F., {Figer}, D.~F., {Hillier}, D.~J., {Geballe}, T.~R., \&
  {Kudritzki}, R.~P. 2009, \apj, 691, 1816

\bibitem[{{Najarro} {et~al.}(2004){Najarro}, {Figer}, {Hillier}, \&
  {Kudritzki}}]{Najarro+2004}
{Najarro}, F., {Figer}, D.~F., {Hillier}, D.~J., \& {Kudritzki}, R.~P. 2004,
  \apjl, 611, L105

\bibitem[{{Najarro} {et~al.}(1994){Najarro}, {Hillier}, {Kudritzki}, {Krabbe},
  {Genzel}, {Lutz}, {Drapatz}, \& {Geballe}}]{Najarro+1994}
{Najarro}, F., {Hillier}, D.~J., {Kudritzki}, R.~P., {et~al.} 1994, \aap, 285,
  573

\bibitem[{{Najarro} {et~al.}(1997){Najarro}, {Krabbe}, {Genzel}, {Lutz},
  {Kudritzki}, \& {Hillier}}]{Najarro+1997}
{Najarro}, F., {Krabbe}, A., {Genzel}, R., {et~al.} 1997, \aap, 325, 700

\bibitem[{{Nugis} {et~al.}(1998){Nugis}, {Crowther}, \& {Willis}}]{Nugis+1998}
{Nugis}, T., {Crowther}, P.~A., \& {Willis}, A.~J. 1998, \aap, 333, 956

\bibitem[{{Nugis} \& {Lamers}(2000)}]{Nugis-Lamers2000}
{Nugis}, T. \& {Lamers}, H.~J.~G.~L.~M. 2000, A\&A, 360, 227

\bibitem[{{Oskinova}(2005)}]{Oskinova2005}
{Oskinova}, L.~M. 2005, \mnras, 361, 679

\bibitem[{{Paumard} {et~al.}(2006){Paumard}, {Genzel}, {Martins}, {Nayakshin},
  {Beloborodov}, {Levin}, {Trippe}, {Eisenhauer}, {Ott}, {Gillessen}, {Abuter},
  {Cuadra}, {Alexander}, \& {Sternberg}}]{Paumard+2006}
{Paumard}, T., {Genzel}, R., {Martins}, F., {et~al.} 2006, \apj, 643, 1011

\bibitem[{{Puls} {et~al.}(2006){Puls}, {Markova}, {Scuderi}, {Stanghellini},
  {Taranova}, {Burnley}, \& {Howarth}}]{Puls+2006}
{Puls}, J., {Markova}, N., {Scuderi}, S., {et~al.} 2006, \aap, 454, 625

\bibitem[{{Reid}(1993)}]{Reid1993}
{Reid}, M.~J. 1993, \araa, 31, 345

\bibitem[{{Schmutz} {et~al.}(1989){Schmutz}, {Hamann}, \&
  {Wessolowski}}]{Schmutz+1989}
{Schmutz}, W., {Hamann}, W.-R., \& {Wessolowski}, U. 1989, \aap, 210, 236

\bibitem[{{Skrutskie} {et~al.}(2006){Skrutskie}, {Cutri}, {Stiening},
  {Weinberg}, {Schneider}, {Carpenter}, {Beichman}, {Capps}, {Chester},
  {Elias}, {Huchra}, {Liebert}, {Lonsdale}, {Monet}, {Price}, {Seitzer},
  {Jarrett}, {Kirkpatrick}, {Gizis}, {Howard}, {Evans}, {Fowler}, {Fullmer},
  {Hurt}, {Light}, {Kopan}, {Marsh}, {McCallon}, {Tam}, {Van Dyk}, \&
  {Wheelock}}]{2MASS}
{Skrutskie}, M.~F., {Cutri}, R.~M., {Stiening}, R., {et~al.} 2006, \aj, 131,
  1163

\bibitem[{{Smith}(1968)}]{Smith1968}
{Smith}, L.~F. 1968, \mnras, 140, 409

\bibitem[{{Smith} \& {Conti}(2008)}]{Smith-Conti2008}
{Smith}, N. \& {Conti}, P.~S. 2008, \apj, 679, 1467

\bibitem[{{Tuthill} {et~al.}(2006){Tuthill}, {Monnier}, {Tanner}, {Figer},
  {Ghez}, \& {Danchi}}]{Tuthill+2006}
{Tuthill}, P., {Monnier}, J., {Tanner}, A., {et~al.} 2006, Science, 313, 935

\bibitem[{{van der Hucht}(2001)}]{vdH2001}
{van der Hucht}, K.~A. 2001, VizieR Online Data Catalog, 3215, 0

\bibitem[{{Wang} {et~al.}(2006){Wang}, {Dong}, \& {Lang}}]{Wang+2006}
{Wang}, Q.~D., {Dong}, H., \& {Lang}, C. 2006, \mnras, 371, 38

\bibitem[{{Wright} \& {Barlow}(1975)}]{Wright+Barlow1975}
{Wright}, A.~E. \& {Barlow}, M.~J. 1975, \mnras, 170, 41

\end{thebibliography}

\end{document}